# Nonlinear Raman Shift Induced by Exciton-to-Trion Transformation in Suspended Trilayer MoS$_2$

Hossein Taghinejad [1], Mohammad Taghinejad [1], Alexey Tarasov [2], Meng-Yen Tsai [2], Amir H. Hosseinnia [1], Philip M. Campbell [2], Ali A. Eftekhar [1], Eric M. Vogel [2], Ali Adibi [1,*]

1: School of Electrical and Computer Engineering, Georgia Institute of Technology, 778 Atlantic Drive NW, Atlanta, GA 30332, USA
2: School of Materials Science and Engineering, Georgia Institute of Technology, 771 Ferst Drive NW, Atlanta, GA 30332, USA

*: Corresponding Author: ali.adibi@ece.gatech.edu

*Abstract:* Layered two-dimensional (2D) semiconductors such as molybdenum disulfide (MoS$_2$) have recently attracted remarkable attention because of their unique physical properties. Here, we use photoluminescence (PL) and Raman spectroscopy to study the formation of the so-called trions in a synthesized freestanding trilayer MoS$_2$. A trion is a charged quasi-particle formed by adding one electron or hole to a neutral exciton (a bound electron-hole pair). We demonstrate accurate control over the transformation of excitons to trions by tuning the power of the optical pump (laser). Increasing the power of the excitation laser beyond a certain threshold (~ 4 mW) allows modulation of trion-to-exciton PL intensity ratio as well as the spectral linewidth of both trions and excitons. Via a systematic and complementary Raman analysis we disclose a strong coupling between laser induced exciton-to-trion transformation and the characteristic phononic vibrations of MoS$_2$. The onset of such an optical transformation corresponds to the onset of a previously unknown nonlinear Raman shift of the in-plane ($E^1_{2g}$) and out-of-plane ($A_{1g}$) vibrational modes. This coupling directly affects the well-known linear red-shift of the $A_{1g}$ and $E^1_{2g}$ vibrations due to heating at low laser powers, and changes it to a nonlinear and non-monotonic dependence with a blue-shift in the high laser power regime. Local reduction of the electron density upon exciton-to-trion transformation is found to be the underlying mechanism for the blue-shift at high laser powers. Our findings enrich our knowledge about the strong coupling of photonic and phononic properties in 2D semiconductors, and enable reliable interpretation of PL and Raman spectra in the high laser power regimes.

Two-dimensional (2D) atomic crystals of layered transition-metal dichalcogenides (TMDCs), such as molybdenum disulfide ($MoS_2$), are promising for a variety of applications complementary to those of graphene. Specifically, the use of TMDCs in field-effect transistors (FETs) with high on/off current ratio $> 10^8$ [1], light emitting devices [2], and efficient control of valley and spin via circularly polarized light [3][4] have been demonstrated. In the two-dimensional limit, reduced dielectric screening alongside the heavy effective mass of charged particles (arising from d-orbitals of Mo atoms [5][6]) significantly enhances Coulombic interaction between electrons and holes. Thus, electrons and holes are tightly bound together and form neutral quasi-particles called excitons. Neutral excitons can further bind to an extra electron (hole) to form negatively (positively) charged excitons, the so called trions [7][8][9]. It has been reported that the formation of such complex quasi-particles influences optical and electrical properties of the layered TMDCs including light absorption and emission [10][11], electron mobility [12][13], and resistance of the metal contacts to $MoS_2$ [14]. The dielectric constant of the media surrounding $MoS_2$ is a critical factor determining not only the binding energies of these quasi-particles but also the relative density of charged and neutral excitons [15][16]. More recently, a binding energy shift of up to 40 meV (for both excitons and trions) and modification of the exciton/trion PL intensity ratio by one order of magnitude has been reported via altering the dielectric constant of the surrounding medium [15]. However, the dynamic control of charge states of these quasi-particles in TMDCs has been only possible so far by applying electrostatic fields in a back-gated FET configuration [7][8]. This usually comes at the cost of unavoidable change of the binding energy of the involved quasi-particles.

Here, we introduce tuning power of the optical pump as an effective way to manipulate charge states of excitons in suspended trilayer $MoS_2$ samples. PL analysis reveals the capability of this technique for effective modulation of trion/exciton intensity ratio as well as the spectral linewidth of the individual emissions with no sign of dragging at the central wavelength. We find that the trion-to-exciton intensity ratio has a parabolic dependence on power of the optical pump while the emission linewidths of both excitons and trions have linear dependence. Moreover, using complementary PL and Raman studies, we reveal a strong coupling between photonic properties and phononic vibrations of the suspended trilayer $MoS_2$ at high laser power regime, induced by exciton-to-trion transformation. Raman analysis reveals that such coupling significantly changes the well-known linear red-shift of the in-plane ($E^1_{2g}$) and out-of-plane ($A_{1g}$) phononic vibrations of $MoS_2$ at low laser power, and turns them into a nonlinear and non-monotonic power dependency at high laser powers. Our experiments suggest that competition between the laser-induced local heating (source of red-shift) and local modulation of electron density induced by exciton-to-trion transformation (source of blue-shift), is the underlying mechanism for this nonlinear effect. To our knowledge, such coupling between the photonic properties and the vibrational modes of layered TMDCs has neither been observed nor experimentally confirmed yet.

High-quality MoS$_2$ films were grown by direct sulfurization of a Mo thin film on a SiO$_2$ substrate. Briefly, a 1nm Mo thin film was deposited on the surface of a silicon wafer covered with 260 nm thick SiO$_2$ layer by e-beam evaporation. Then, the sample was placed into a furnace and exposed to a controlled flow of sulfur gas at a temperature of 1050 $^0$C to grow MoS$_2$. Details of the growth process are reported elsewhere [17]. The MoS$_2$ film was then transferred onto a new substrate using a wet-transfer technique described in the experimental section. The new substrate has a 320 nm thick SiO$_2$ layer thermally grown on Si wafer and patterned with an array of holes with diameters of 6-7 μm. Figure 1a shows an optical microscope image of the transferred trilayer MoS$_2$ on the patterned substrate and the arrow points to a typical region where MoS$_2$ is suspended over the holes. All the following measurements were performed on these suspended zones which allow study of the intrinsic properties of MoS$_2$ without undesirable contributions from the substrate.

Figure 1b illustrates the PL spectra of the suspended trilayer MoS$_2$ carried out at different laser powers. All the PL spectra were recorded at room temperature in the reflection mode with laser excitation wavelength of 488 nm. Using an X50 objective lens, the laser was focused at the center of the hole away from edges with the diameter of the spot kept at ~ 1 μm. Two sets of emission peaks can be observed in these spectra (Figure 1b) at different laser powers. These are attributed to the direct transition at the K-point from the bottom of the conduction band to the top of the valence band (the A exciton), and from the bottom of the conduction band to the split up band in the valence band (the B exciton) [18][19][20]. The A exciton emission is further divided into two sub-bands associated with neutral A excitons, A$^0$, and the negatively charged excitons, A$^-$, the so called trions [7][21]. As can be inferred from Figure 1b, the strength of the splitting depends on the power of the excitation pump, and the relative contribution of the $A^0$ and $A^-$ emissions to the whole spectrum varies with the pump power.

In order to investigate this effect more precisely, deconvolution of the PL spectra at different laser powers is performed and shown in Figures 2a-f. In these plots, black lines and red lines represent the experimental data and the Lorentzian fits, respectively. The fitting is performed assuming the PL intensity to be the sum of three individual Lorentizan lineshapes (blue lines) associated with the B excitons, the neutral A excitons ($A^0$), and the negatively charged A excitons (trions, $A^-$), as labeled in Figure 2a. The power dependent splitting of the A exciton feature into two sub-bands can be inferred clearly from the sequence of the PL spectra in Figures 2a-f. It is clear that the contribution of the neutral excitons, $A^0$, becomes weaker at higher laser powers and the emission from the trions, $A^-$, takes over almost the entire spectrum at higher powers (inserted arrows highlight eminence of the $A^-$ emission as power increases). The observation of the $A^-$ peak at 1 mW laser power (Figure 2a) points to the n-type doping of the $MoS_2$ layers, consistent with our previous electrical characterization of FET devices made from similarly grown trilayer $MoS_2$ films [17] [22]. A similar spontaneous n-type doping of $MoS_2$ has also been reported in few-layer $MoS_2$ exfoliated from bulk crystals [1][7][23]. No evidence of significant wavelength modulation was observed for neither of the $A^0$ or $A^-$ features induced by laser at different powers.

To gain more insight into the laser-induced exciton-to-trion transformation effect, quantitative interpretation of the PL spectra is presented in Figure 3. As seen in Figure 3a, both excitons and trions have stronger PL intensities at higher laser powers. However, unlike the trions, excitons experience intensity saturation for laser powers higher than above 4-5 mW. It implies that the trion-to-exciton intensity ratio, i.e. $I(A^-)/I(A^0)$, has a strong laser power dependence. Figure 3b depicts the variation of I(A⁻)/I(A⁰) ratio with the laser power which clearly shows that at lower laser powers the I(A⁻)/I(A⁰) remains fairly constant while it rises up rapidly at high laser powers. This variation can be well fitted with a parabolic function of the form $ap^2 + b$ where $p$ is the laser power in mW, $a = 2.49 \times 10^{-2}$ (mW)$^{-2}$ is a nonlinear fitting parameter describing the strength of the exciton-to-trion transformation, and $b = 2.78$ is a dimensionless parameter. Indeed, the parameter '$b$' accounts for the spontaneous doping of the pristine MoS$_2$ sample [7], since otherwise the intensity of the trion emission (and hence I(A⁻)/I(A⁰)) would be negligible at low laser powers. Direct modulation of I(A⁻)/I(A⁰) by controlling the laser power can be clearly observed in Figure 3b. Consistent with Figure 3a, significant modulation of this intensity ratio starts above the threshold power of ~ 4 mW while the modulation below this value is negligible. We will later see that such a threshold power can be traced via Raman spectroscopy of freestanding MoS$_2$ as well. From the standpoint of the PL linewidth (Figure 3c), increasing the power linearly broadens the trion emission while linearly narrowing the exciton line. Similar to the intensity modulation, impact of the laser power on the trion linewidth is more pronounced, as inferred from the larger slope of the trion linewidth compared to that of the exciton linewidth (~ 2 times higher, dashed lines in Figure 3c). However, due to more phonon scattering at room temperature the linewidths of both A⁰ and A⁻ are broadened compared to previously reported values measured at very low temperatures (20-100 K) [7][8].

Working in the high laser power regime demands precise consideration of laser induced local heating of the freestanding MoS$_2$. Since the potential energy of the MoS$_2$ lattice has anharmonic terms [24], it has been demonstrated that heating through the thermal population factors of interacting phonons can affect vibrational properties of this material [25]. In this regard, Raman spectroscopy via monitoring the peak shifts of the in-plane ($E^1_{2g}$) and out-of-plane ($A_{1g}$) vibration modes of MoS$_2$ has been extensively used [24][26][27]. These studies have focused on the low power regime (< 1mW), and used supported or suspended exfoliated MoS$_2$ flakes [28]. However, the effect of high laser power excitation on vibrational modes of suspended MoS$_2$ samples has not been explored.

Figure 4a illustrates power dependent Raman spectra collected from a freestanding trilayer MoS$_2$ sample with $A_{1g}$ and $E^1_{2g}$ vibrations labeled on the plot. For the sake of comparison, Raman spectra were carried out at the same excitation wavelength and laser powers as those of the PL measurements presented above. Figures 4b-c depict the peak positions of the $A_{1g}$ and $E^1_{2g}$ vibration modes measured at different laser powers. Interestingly, the experimental data (red squares) reveals a nonlinear and, more importantly, non-monotonic shift of the $A_{1g}$ and $E^1_{2g}$ peak positions in response to increasing the laser power. Both features initially undergo a red-shift at low laser powers which later turns into a blue-shift at higher powers. Such non-monotonic behavior cannot be completely explained by a simple laser-induced local heating effect. Although such heating can account for the red-shift at low laser powers [26][27], it cannot explain the blue-shift at higher laser powers. To explore this effect in more detail, parabolic functions of the form $\alpha_{1,2} p^2 + \beta_{1,2} p + \gamma_{1,2}$ were fitted to the experimental data (broken lines in Figures 4b-c) where $p$ is the power in mW and $\alpha_{1,2}$, $\beta_{1,2}$ and $\gamma_{1,2}$ are the fitting parameters with subscript '1' and '2' are used for $A_{1g}$ and $E^1_{2g}$, respectively. Numerical values of these parameters, shown in Figures 4b-c, contain two important observations: 1) the linear coefficients $\beta_1$ and $\beta_2$ are negative while the nonlinear coefficients $\alpha_1$ and $\alpha_2$ are positive; 2) the linear coefficients, i.e. $\beta_1$ = -3.54×10$^{-1}$ ($\frac{cm^{-1}}{mW}$) and $\beta_2$ = -2.89×10$^{-1}$ ($\frac{cm^{-1}}{mW}$), are approximately one order of magnitude larger than the nonlinear coefficients, $\alpha_1$ = 4.3×10$^{-2}$ ($\frac{cm^{-1}}{mW^2}$) and $\alpha_2$ = 3.83×10$^{-2}$

($\frac{cm^{-1}}{mW^2}$). Therefore, at low laser powers, the contribution of the nonlinear term is negligible, and the shift of the peak positions depends linearly on the power, with negative slopes ($\beta_{1,2}$) accounting for the red-shift. Our description agrees well with the previous explanation of such red-shift based on the thermal heating produced by the laser exposure [24][26][27]. However, when the laser power is sufficiently high, nonlinear terms dominate and the peak position shifts become proportional to the square of the laser power with positive proportionality constants $\alpha_1$ and $\alpha_2$ representing the blue-shift (Figures 4b-c). Considering the specific laser power at which the minimum of the parabola occurs provides insightful information about the nature of such non-monotonic dependence. These values, found by setting the derivative of the fitting functions to zero, are $P_1$ = 4.1 mW and $P_2$ = 3.77 mW for $A_{1g}$ and $E^1_{2g}$ modes, respectively. Importantly, these values are essentially the same as the power at which significant exciton-to-trion transformation initiates, as we discussed in the PL analysis of Figure 2 and Figures 3 a-b. Indeed, the good agreement between these critical powers in the PL and Raman spectra suggests that the effective exciton-to-trion transformation at high laser powers is responsible for the observation of nonlinear and non-monotonic power dependent Raman shifts in the freestanding trilayer $MoS_2$. Also, comparing the fitted numerical values of the nonlinear coefficients describing the Raman shifts of the $A_{1g}$ and $E^1_{2g}$ features (i.e. $\alpha_1$ = 4.3×10$^{-2}$ ($\frac{cm^{-1}}{mW^2}$) and $\alpha_2$ = 3.83×10$^{-2}$ ($\frac{cm^{-1}}{mW^2}$)) with the nonlinear coefficient describing exciton-to-trion transformation (i.e. $a = 2.49 \times 10^{-2}$ mW$^{-2}$) demonstrates similar levels of power dependence. In other words, a nonlinearly power-dependent and optically induced exciton-to-trion transformation in $MoS_2$ is translated, with similar impact, into a nonlinear shift of the vibrational phononic modes in the Raman spectra at high laser powers. The underlying mechanism for this nonlinearity could be attributed to the local modulation of charge carriers upon exposure to the laser light at high powers. Trion formation is possible through addition of an extra electron to the neutral exciton which in turn reduces the local number of the electrons by one. As the power goes up, supported by results of the PL analysis, the rate of such transformation rapidly increases particularly at laser powers above a threshold value of 4 mW in the case of our experiments. The consumption of electrons due to the formation of trions substantially reduces local density of electrons in the exposed area. On the other hand, it has been shown that the electron density reduction produces a blue-shift in the Raman spectrum of $MoS_2$ due to less effective electron-phonon interaction [29][30]. Therefore, the effect of local modulation of the electron density during exciton-to-trion transformation becomes comparable to and even exceeds the local heating effect at higher laser powers (e.g. > 4 mW in our sample) and changes the Raman peak shift direction from red to blue. Also, group theory analysis predicts that the symmetric $A_{1g}$ vibrations couple more strongly to electrons compared to the $E^1_{2g}$ mode [29] and hence, the former peak is more sensitive to the change of the laser power. We also observe this effect by noting that the parameter $α_1$ is larger than $α_2$.

For the sake of completeness, the effect of laser power on the intensity of $E^1_{2g}$ and $A_{1g}$ modes is also shown in Figure 4d. As seen, the intensities of both $E^1_{2g}$ and $A_{1g}$ vibrations scale linearly with the laser power while the slope of the $A_{1g}$ intensity is slightly larger than that of $E^1_{2g}$, most likely caused by the higher degree of freedom the out-of-plane vibrations have [31]. As a direct consequence, $A_{1g}/E^1_{2g}$ intensity ratio is nearly independent of the laser power. This value, shown in Figure 4e, is approximately 1.35 for suspended trilayer $MoS_2$ grown by sulfurization of Mo thin films.

In conclusion, we studied the photoluminescence and Raman spectra of synthesized suspended trilayer $MoS_2$ as a function of laser power. We observed a previously unexplored strong coupling between the optical transitions and the phononic modes of $MoS_2$ in the high laser power regime. Controlling power of the optical pump (laser) was introduced as an effective way to change exciton and trion states in freestanding $MoS_2$ samples. As a direct consequence of the exciton-to-trion transformation a nonlinear dependence of the Raman peak shift on the laser power was revealed for both $A_{1g}$ and $E^1_{2g}$ modes. We also showed that while laser induced local heating effects dominate the Raman shifts and leads to a red-shift at low laser powers, the increased trion formation overcompensates the heating and causes a blue-shift at high laser powers (> 4 mW). The underlying mechanism for the Raman blue-shift at high laser powers was found to be the local reduction of electron density, caused by trion formation. These results are important for the fundamental understanding of the interplay between optical transformations and phonon modes in 2D materials, and they allow reliable interpretation of PL and Raman data in the high laser power regime.

*Acknowledgement:* This work was primarily funded by the Air Force Office of Scientific Research (AFOSR) under Grant No. FA9550-13-1-0032 (G. Pomrenke). The material development part is also supported by: the Swiss National Science Foundation (SNF) within the Early Postdoc Mobility Program under Grant No. P2BSP2_148636 (to A.T.); the National Science Foundation (through CBET-1264705); the Center for Low Energy Systems Technology, one of six centers supported by the STARnet phase of the Focus Center Research Program, a Semiconductor Research Corporation program sponsored by MARCO and DARPA; and the Georgia Tech Research Institute Robert G. Shackelford Fellowship ( to P.M.C.).

Experimental Section:

- *MoS$_2$ Transfer:* After MoS$_2$ growth on SiO$_2$, a thick layer of Poly-methyl-methacrylate (PMMA) was spin-coated on the surface of the sample. Then, the sample was dipped into buffered oxide etch overnight to etch the 260 nm supporting SiO$_2$ layer. The MoS$_2$ was released from the substrate and was floating on the liquid, supported by the PMMA. Then, floating MoS$_2$/PMMA was fished with the target substrate containing patterned hole arrays. Finally, PMMA was removed from the MoS$_2$ surface by immersion in acetone for 2 hours and rinsing with deionized water.

- *Photoluminescence (PL) measurements:* PL measurements were carried out in a reflection mode with excitation laser wavelength of 488 nm at different laser powers of 1 mW, 2.5 mW, 4 mW, 6.5 mW, 8 mW, and 10 mW. The system was equipped with a single-pass spectrometer with a grating of 1800 grooves/mm and a Peltier-cooled CCD array to minimize effects of thermal variations on measurements. Using a 50X objective lens light was focused at the center of the hole (with a laser spot diameter of 1 μm).

- *Raman Measurements:* Raman spectra were collected in a dispersive micro-Raman spectrometer (Thermo Nicolet Almega XR) in back scattering configuration with excitation wavelength of 488 nm (Ar laser). The laser power was set to 1 mW, 2.5 mW, 4 mW, 6.5 mW, 8 mW and 10 mW. 50/50 beam splitters and notch filters were centered at the laser line to reject the Rayleigh scattering. The laser beam was focused at the center of the holes using a 50X objective lens and the Raman scattered light was collected with the same objective lens. Diameter of the laser spot size at the MoS$_2$ sample was kept fixed at 1μm, small enough compared to the hole diameters (6μm-7μm).

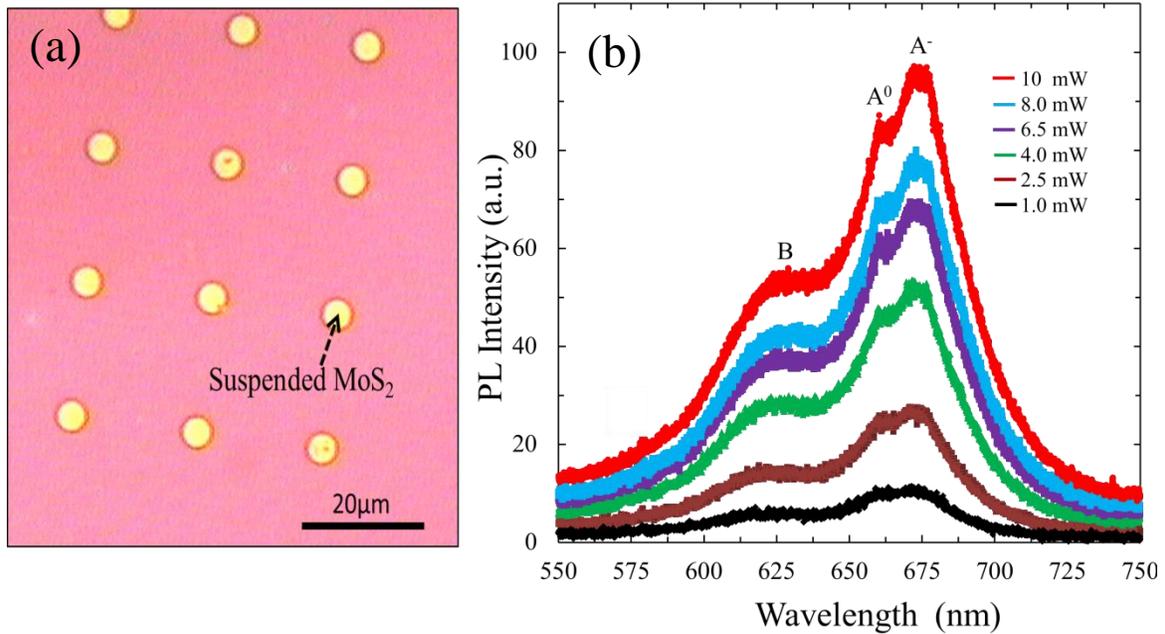

Figure 1: (a) Optical microscopy image of the trilayer $MoS_2$ on 320 nm patterned $SiO_2$ substrate containing an array of holes of 6-7 μm diameters. A hole is indicated by an arrow, on top of which the $MoS_2$ layer is suspended. (b) Photoluminescence (PL) spectra of the suspended $MoS_2$ collected at different laser powers ranging from 1 mW to 10 mW. Three different emission features can be identified in these spectra: B excitons, neutral A excitons ($A^0$), and negatively charged A excitons ($A^-$), the so-called trions. Power dependent splitting of the A excitons can be readily inferred from these spectra. Excitation laser has wavelength of 488 nm with a spot diameter of 1μm focused at the center of the holes. All the measurements have been carried out at room temperature.

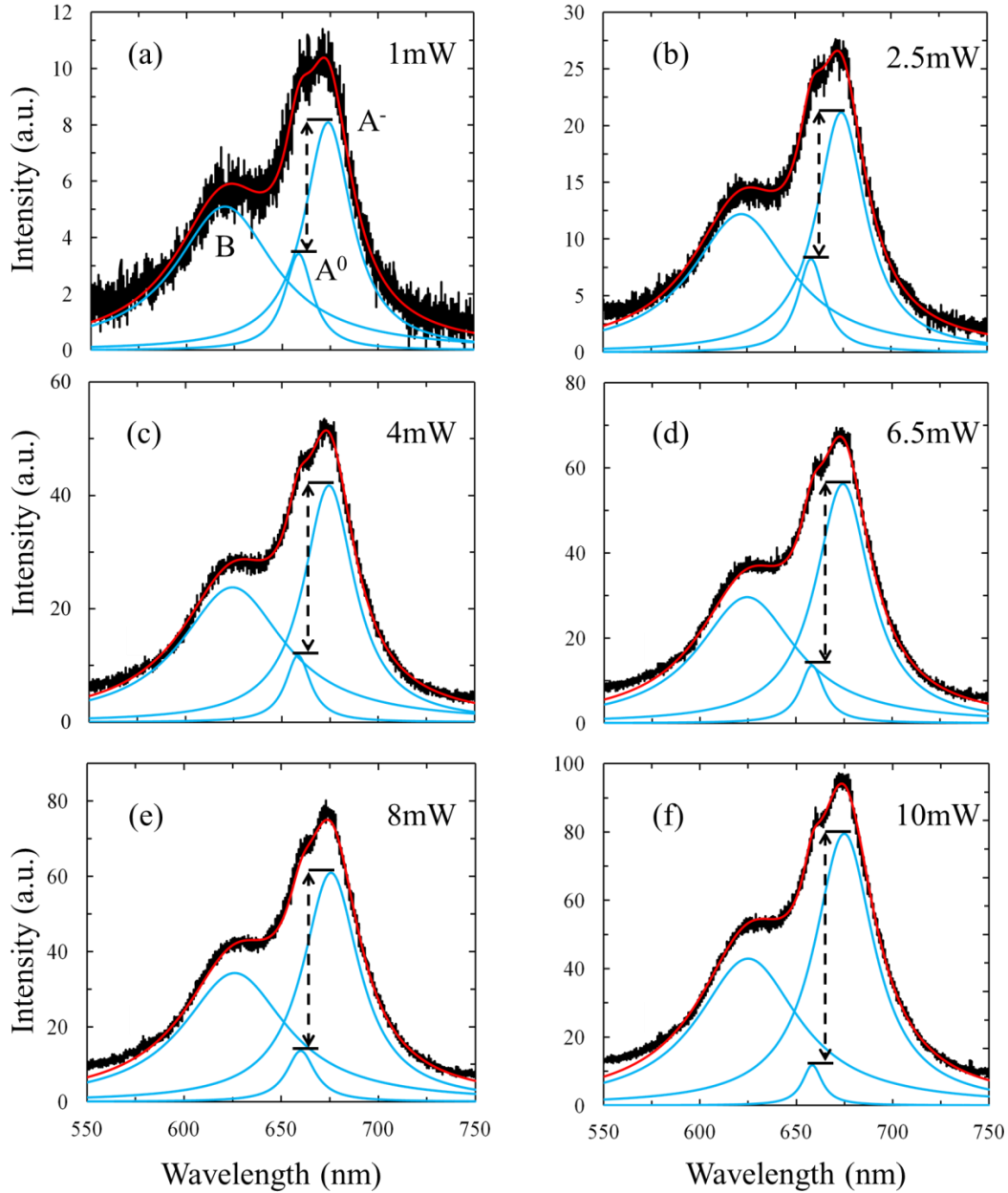

**Figure 2:** Deconvolution of photoluminescent spectra of the suspended trilayer $MoS_2$ recorded at different excitation laser powers: (a) 1 mW, (b) 2.5 mW, (c) 4 mW, (d) 6.5 mW, (e) 8 mW, and (f) 10 mW. Black lines represent experimental data and the red lines are Lorentzian fits. Blue lines are components of the Lorentzian fits representative of luminescence from B excitons, neutral A excitons ($A^0$), and negatively charged A excitons ($A^-$). As indicated by arrows, an overall suppression in relative contribution of the $A^0$ emission is observed as the power of the excitation laser increases while the trion emission, $A^-$, becomes more significant.

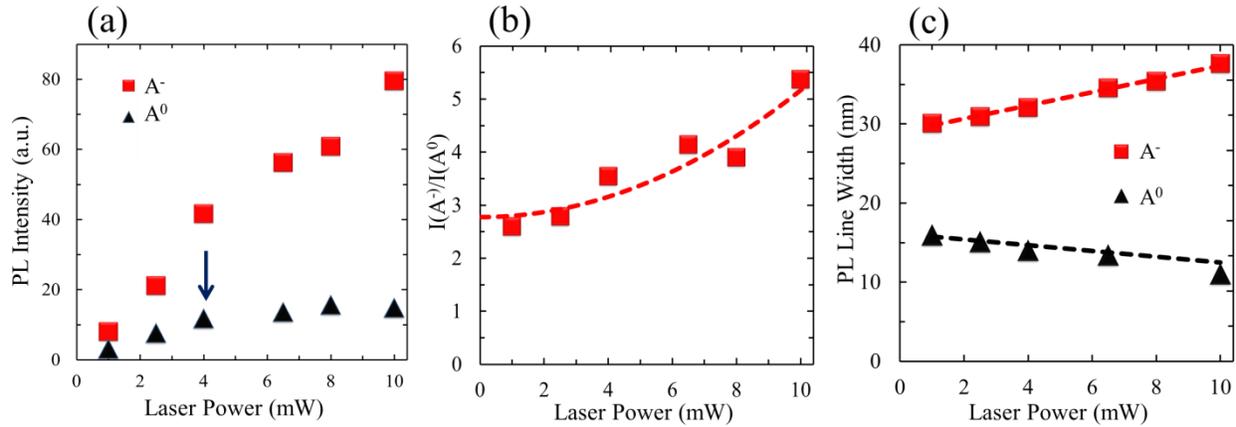

**Figure 3:** (a) Photoluminescence intensity of trion, A⁻, and exciton, A⁰, at different laser powers. Saturation of the exciton intensity can be observed at high laser powers (> 4 mW). The arrow points to the beginning of exciton intensity saturation. (b) Trion-to-exciton intensity ratio, I(A⁻)/I(A⁰) showing a nonlinear power dependence: squares are the experimental values and the broken line is a parabolic fit of the form $ap^2 + b$ with $p$ being the laser power, $a = 2.49 \times 10^{-2}$ (mW)$^{-2}$, and the dimensionless constant $b = 2.78$. (c) Linewidths of the exciton and trion as a function of the laser power. Trion linewidth broadens while the exciton linewidth becomes narrower; both change linearly with the laser power. Dashed lines are linear fits to the experimental data.

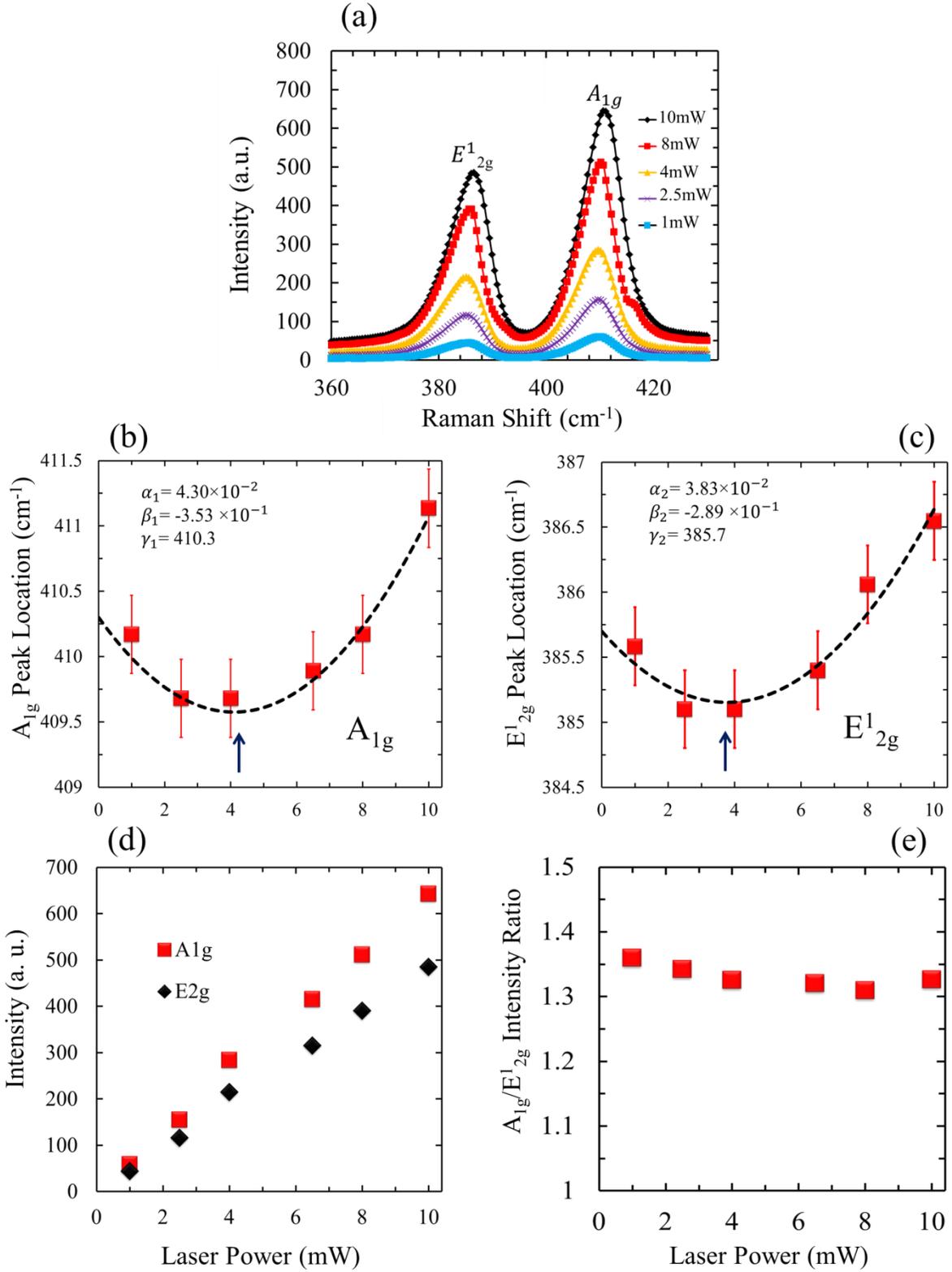

**Figure 4:** (a) Effect of laser power on characteristic Raman spectra of freestanding trilayer MoS$_2$. E$^1_{2g}$ and A$_{1g}$ modes are labeled on the plot. The power-dependent variation of the positions of the peaks of the $A_{1g}$ (b) and $E^1_{2g}$ (c) vibrations represent a nonlinear and non-monotonic dependence on laser power starting with a red-shift at low powers that turns into a blue-shift at higher powers. Squares are experimental data and dashed lines are parabolic fits of the form $\alpha p^2 + \beta p + \gamma$, where $p$ is the laser power and $\alpha$, $\beta$, $\gamma$ are the fitting parameters. Numerical values of the fitting parameters are illustrated with subscript 1 for $A_{1g}$ and 2 for $E^1_{2g}$ on the associated plots. For both features, the effect of the laser-induced local carrier modulation ($\alpha p^2$) becomes comparable to the local heating effect ($\beta p$) at ~ 4 mW laser power. Arrows point to the critical power at which the red-shift turns into a blue-shift. (d) Peak intensities of both $E^1_{2g}$ and $A_{1g}$ vibrations increase linearly with laser power while (e) $A_{1g}/E^1_{2g}$ intensity ratio remains approximately constant at ~ 1.35. The wavelength of the laser is 488 nm with ~ 1 µm diameter spot focused at the center of 6-7µm holes (see Figure 1a). Error bars are 0.3 cm$^{-1}$, equal to the resolution of the Raman spectrometer.